\documentclass[namedreferences]{solarphysics}
\usepackage[optionalrh,solaromanenum]{spr-sola-addons} 
\usepackage{graphicx}                    
\usepackage{amssymb}                    
\usepackage{color}                       
\usepackage{url}                         

\begin{document}

\begin{article}

\begin{opening}

\title{Kinematics of ICMEs/shocks: blast wave reconstruction using type II emissions}

%
\author{P.~\surname{Corona-Romero}$^{1}$\sep
        J.A.~\surname{Gonzalez-Esparza}$^{1}$\sep
        E.~\surname{Aguilar-Rodriguez}$^{1}$\sep
        V.~\surname{De-la-Luz}$^{1}$\sep
        J.C.~\surname{Mejia-Ambriz}$^{1}$
       }

%
\runningauthor{Corona-Romero et al.}
\runningtitle{Draft}

%
   \institute{$^{1}$ SCiESMEX, Instituto de Geofisica, Unidad Michoacan, Universidad Nacional Autonoma de Mexico, Antigua carretera a Patzcuaro \# 8701 Ex-Hda. San Jose de la Huerta Morelia, Michoacan, C.P.58089, Mexico.\\
                     P. Corona-Romero, email: \url{pcorona@geofisica.unam.mx}\\ 
                     J.A. Gonzalez-Esparza, email: \url{americo@geofisica.unam.mx} \\ 
                     E. Aguilar-Rodriguez, email: \url{ernesto@geofisica.unam.mx}\\
                     V. de-la-Luz, email: \url{vdelaluz@geofisica.unam.mx}
                     J.C. Mejia-Ambriz, email: \url{jcmejia@geofisica.unam.mx}
             }

\begin{abstract}
We present a physical methodology to reconstruct the trajectory of interplanetary shocks using type II radio emission data. This technique calculates the shock trajectory assuming that the disturbance propagates as a blast wave in the interplanetary medium. We applied this Blast Wave Reconstruction (BWR) technique to analyze eight fast Earth-directed ICMEs/shocks associated with type II emissions. The technique deduces a shock trajectory that reproduces the type II frequency drifts, and calculates shock onset speed, shock transit time and shock speed at 1~AU. There were good agreements comparing the BWR results with the type II spectra, with data from coronagraph images,  {\it in situ} measurements, and interplanetary scintillation (IPS) observations. Perturbations on the type II data affect the accuracy of the BWR technique. This methodology could be applied  to track interplanetary shocks causing TII emissions in real-time, to predict the shock arrival time and shock speed at 1~AU. 
\end{abstract}

%

\end{opening}

%
\section{Introduction}
     \label{introduction} 

Coronal mass ejections (CMEs) associated with interplanetary (IP) shocks are the most important phenomena of solar activity for space weather purposes. Together, IP counterparts of CMEs (ICMEs) and shocks are the main triggers of major geomagnetic storms when they interact with the Earth's magnetosphere ({\it e.g.} \opencite{ontiveros2010}). Since fast Earth-directed ICMEs and associated shocks (from now on ICMEs/shocks) represent potential hazards to the geomagnetic stability, tracking them in the IP medium is a priority for space weather purposes. In general, the tracking of an ICME/shock event requires combining multiple observations, by different instruments, in order to approximate their trajectories in the inner heliosphere ({\it e.g.} \opencite{bisi2010}). However, despite the current observational capabilities, such tracking is still a very difficult task.

The kinematics of ICMEs/shocks has a major interest in space weather studies. It is well known that fast CMEs ($>900\,km\,s^{-1}$) drive shock waves near the Sun \cite{vourlidas2003}, and that IP shocks significantly decelerate as they propagate in the IP medium \cite{gosling1968}. Furthermore, comparing near-Sun (plane-of-sky) and {\it in situ} (at 1 AU) speeds, shows that fast CMEs decelerate as they propagate through IP space ({\it e.g.} \opencite{lindsay1999}; \opencite{gopalswamy2000}; \opencite{gopalswamy2005}). In general, the initial speed of fast CMEs as measured in coronagraph images is always faster than its corresponding transit speed at 1~AU, which in turn is faster than its {\it in situ} ICME speed at 1~AU. In fact, \inlinecite{corona2011} argue that this deceleration indicates that IP shocks are not driven any more by their corresponding ICMEs when they reach the Earth's orbit.

The speed of IP shocks also decreases with heliocentric distance (\opencite{dryer1974}; \opencite{pinter1982}; \opencite{gopalswamy2005}; \opencite{reiner2007}) and shock's speed apparently decreases with the inverse of the square-root of distance (\opencite{dryer1984}; \opencite{smart1985}). Such a decaying rule is a signature for blast waves in density profiles that decay with the square of distance, like IP medium does (see, \opencite{rogers1958}; \opencite{cavaliere1976}). Then, the deceleration of IP shocks was interpreted as a blast wave propagation ({\it e.g.}, \opencite{smart1985}; \opencite{pinter1990}) which may start somewhere before 1 AU \cite{corona2011}. In fact, \inlinecite{feng2010} empirically concluded that many of {\it in situ}-detected (at 1 AU) IP shocks already evolve as blast waves. Furthermore, \inlinecite{pinter1990} and \inlinecite{corona2012c} independently found that trajectories of IP shocks are well described by blast wave propagation for heliocentric distances larger than $\sim 25-45\, R_\odot$. Then, for the purposes of this work, we will assume that trajectories of IP shocks can be approximated by blast wave equation.


Type II (TII) decametric-kilometric radio burst emissions is a useful signature to track the propagation of an IP shock \cite{pinter1982,pinter1990}. These events are characterized by a narrow band of intense radiation drifting to lower frequencies as time increases (distance from the Sun). TII emissions are produced by the excitation of plasma waves in the ambient solar wind by a disturbance (say an ICME/shock) propagating outward from the Sun \cite{cane1984,cane1987}. These TII emissions occur at the fundamental and/or harmonic of the plasma frequency ($f$), which is proportional to the square root of the local electron density ($n$), at the source region.

TII emissions usually start at frequencies below $150~{\rm MHz}$, where the disturbance is just a few solar radii away from the Sun, and may extend down to the kilometric domain, slowly drifting to lower frequencies all the way to 1 AU, where the local plasma frequency of the solar wind is about $25~kHz$. Type II emissions are classified according to their wavelength regime as: metric (m), decameter/hectometric (DH), and kilometric (km) bands. It is now well established that DH to km TII emissions are caused by the propagation of  ICMEs/shocks through the IP medium  \cite{cane1984,cane1987}. However, we should keep in mind that not necessarily all ICME/shock events generate Type II radio emissions \cite{gopalswamy2008}.

Since beyond the supermagnetosonic point the solar wind density decays with the square of the heliocentric distance ($n \propto r^{-2}$), the frequency-drifting of TII emissions allows us to track the propagation of their source region. These radio observations can be used to reconstruct the propagation of an ICME/shock. \inlinecite{pinter1982} and \inlinecite{pinter1985} studied the propagation of IP shocks by using several techniques, including frequency drift of the associated TII emissions. Subsequently, \inlinecite{pinter1990} used data derived from TII emissions to develop a semi-empirical model to approximate IP shocks trajectories and transit times from the Sun to Earth. In their model the shock trajectory has two stages: an initial short interval of constant speed, followed by a blast-wave propagation. Subsequently, \inlinecite{watari1998} used this blast wave model to explore relations between transit times and arrival speeds (at 1 AU) of IP shocks.

\inlinecite{leblanc2001} traced shocks/CMEs from the solar corona up to 1 AU, by combining white-light coronagraph, TII and {\it in situ} data. Afterwards, \inlinecite{reiner2007} proposed a kinematic model for the propagation of ICMEs/shocks which reproduces the TII frequency drift. This model assumes an arbitrary kinematics where initially the CME/shock suffers a strong deceleration followed by a constant-speed propagation. \inlinecite{cremades2007} suggested a method to predict the arrival of ICMEs/shocks associated with type II emissions assuming a constant velocity in the IP medium. On the other hand, following the work by \inlinecite{reiner2007}, \inlinecite{gonzalez2009} suggested a technique to approach the local shock speed using TII emissions. This technique assumes that the shock speed can be considered  constant during short time intervals. By analyzing a few case events combining coronagraph, IPS, {\it in situ} and TII data, they found that fast shocks/ICMEs gradually decelerate from near the Sun to 1 AU.


%
The aim of this paper is to describe a new physical reconstruction of ICME/ shock propagation using TII data. The methodology assumes that IP shocks decelerate as blast-waves and uses TII data to reconstruct the shock trajectory from near the Sun ($\sim 6 \, R_\odot$) to 1~AU. This reconstruction deduces: (1) the shock initial speed (which can be compared with the CME initial speed reported by SOHO/LASCO), (2) shock arrival speed and (3) shock transit time to 1~AU (which both can be compared with {\it in situ} measurements). We analyzed eight case events of fast-halo Earth-directed CMEs associated with kilometric TII emissions and ICME/shock counterparts. In general, we found a very good agreement comparing the results obtained by the blast wave reconstruction technique and the observations. The outline of the paper is as follows: Section~\ref{modelo} explains the reconstruction technique using TII emissions; Section~\ref{results} presents the analysis of eight case events; and Sections~\ref{discussion} and \ref{conclusions} present the discussion and conclusions respectively.

\section{Physical reconstruction of shock trajectories using Type II data}
 \label{modelo}

The methodology to obtain the trajectory of ICME/shock events associated with TII emissions can be summarized in three steps as follows: (1) from the TII data we obtain the central emission frequencies and their associated bandwidths; (2) we apply a blast wave reconstruction model to the shock propagation using TII data; (3) we compare directly the shock parameters obtained by the reconstruction with chronagraph and {\it in situ} measurements.

\subsection{Type II central emission frequency}
 \label{technique2}
 \begin{figure} 
 \centerline{\includegraphics[width=0.7\textwidth,clip=]{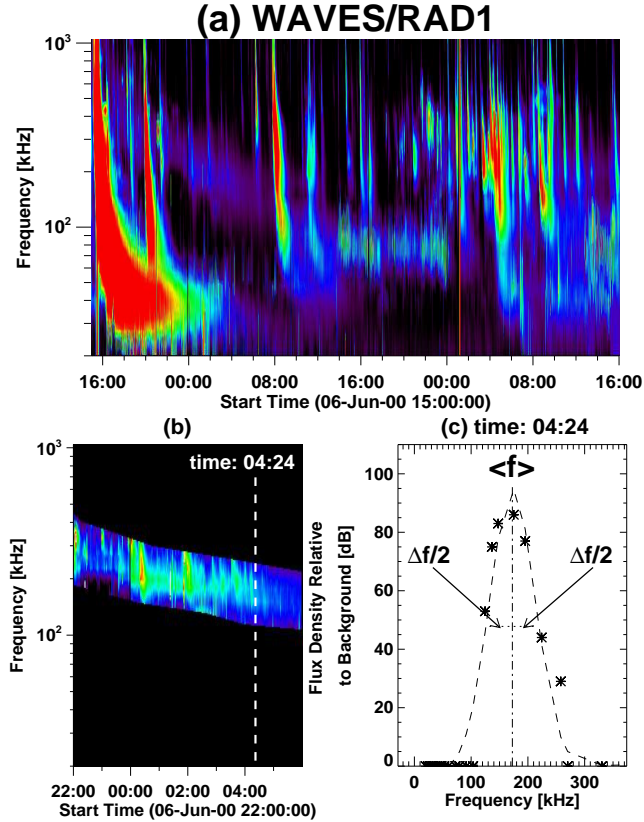}}
 \caption{Case study, central frequency and associated bandwidth deduction for event 1 (LASCO/CME on 6 June 2000). (a) WAVES/RAD1 dynamic spectrum associated with the propagation of the ICME/shock. (b) Isolated TII emission between 6-Jun-2000 22:00 UT to 7-Jun-2000 06:00 UT. (c) Flux density {\it vs} frequency for 1-min RAD1 spectrum at 04:24 UT. White dashed-line in panel (b) marks a temporal sample (at 04:24 UT) to perform the frequency analysis and the dashed black-line, in panel (c) points out the gaussian fit. Further description of the figure in the text.}\label{figure0}
 \end{figure}
In practice, most observations of TII radio emissions are complex and show important fluctuations, and even gaps, in intensity, frequency  emission and drift rate. Most of these fluctuations derive from the solar wind structure, the IP magnetic field topology and the shock front evolution \cite{knock2005}. Due to such complexity, one of the problems in analysing TII emission is to identify the central emission frequency ($\langle f \rangle$) of the TII radio burst and the associated bandwidth ($\Delta f$). In order to simplify this task, we apply the technique by \inlinecite{aguilar2005} and \inlinecite{gonzalez2009} to infer $\langle f \rangle$ and its associated $\Delta f$. We present in Figure \ref{figure0} a case study to illustrate this technique. Figure~\ref{figure0}a shows the radio spectrum of event 1 (see Table \ref{tabla_eventos}) detected by Radio Receiver Band 1 (RAD1) \url{http://lep694.gsfc.nasa.gov/waves/data_products.html}. RAD1 is one of three radio receivers of the {\it WAVES} experiment on board the {\it WIND} spacecraft \cite{bougeret1995}. RAD1 covers a frequency range from 20 to 1,040 kHz in a frequency band divided into 32 real channels. The frequency emission is observed at 15:00 UT on 6 June at 600 kHz, drifting slowly down to 60 kHz at 00:30 UT on 8 June. The TII emission is characterized by a single tone with intervals of very intense emission, which lasts several hours. 

In Figure~\ref{figure0}b we isolated the TII event from the dynamic spectrum by setting to zero any emission outside the TII emission feature. Figure~\ref{figure0}b also shows a vertical dashed-line indicating a time sample (04:24 UT) on the TII emission. Figure~\ref{figure0}c shows this sample as a flux density {\it vs} frequency plot (asterisks), it also presents the Gaussian fit (dashed-line) to the flux-density distribution. This Gaussian approximation computes the values for $\langle f \rangle _i$ (dash-dotted line) and $\Delta f_i/2$ (dotted-lines) at the given time ($t_i$) (where $i$ is the index in the data series). The systematic repetition of this process, at every time sample of the spectrum, gives us also criteria to determine whether the time sample in the spectrum has a well-defined central emission or not ({\it e.g.}, there are some spectra with multiple maxima in the flux density distribution, for these cases we discard the data point). After applying this filtering technique to the TII data, we obtain a set of $m$ values for $\langle f \rangle _i$, $\Delta f_i/2$ and $t_i$, with $1 \leq i \leq m$. 

\inlinecite{corona2012c} simulated TII emissions using a physical analytical model of ICMEs/shocks propagation. These synthetic radio bursts did not follow the central frequency of the associated TII emission but their lower frequency edge ($\underline{f}$). Assuming that the shock front is the TII source emission, the shock leading front is at the farthest distance from the Sun, which is associated with lower solar wind densities emitting at lower frequencies. Based on this previous experience, we will use the lower frequency edge ($\underline{f}$) of the TII emission for the regression model. From the analysis of radio data commented on before (see Figure \ref{figure0}), we define the lower frequency edge as $\underline{f}_i = \langle f \rangle _i - \Delta f_i/2$. As we will explain below, with these frequency data we can reconstruct the trajectory of the IP shock.

\subsection{Blast Wave Regression technique}
 \label{technique2}
 
The Blast Wave Regression (BWR) technique reconstructs the trajectory of shock fronts using TII data. In order to do so, this technique combines TII frequency data with the kinematic equations of a blast wave. A blast wave is a shock wave which has a finite amount of energy, the shock is transferring energy to the medium, decaying continuously as it propagates. In the case of an IP shock, a blast wave propagation implies that the shock decelerates following a particular rate  \cite{rogers1957}. On the other hand, a driven shock is a shock wave that has a continuous source of energy from a driver, this source input energy replaces the energy that the shock transfers to the medium. In the case of an IP shock, a driven shock propagation implies that the shock and its driver (ICME) propagate at almost at almost the same speed \cite{parker1961}. Previous studies,  using {\it in situ} data, suggest that IP shocks are not driven by their ICMEs further away from 1~AU \cite{burlaga1981,feng2010}. In fact, \inlinecite{corona2011} claim that the CME/shock driven stage lasts just a few hours from which the IP shock propagation follows a blast wave decay. We should keep in mind, as commented on before, that the deceleration of fast CMEs/shocks in the IP medium is well established. For this study, we assume that IP shock waves follow a blast wave propagation  from a certain heliocentric distance, which on average is about 15~R$_{\odot}$ (this value is obtained from the reconstruction algorithm).

\inlinecite{cavaliere1976} investigated the self-similar solutions for shock waves propagating through different density profiles. According to them, when density decays as $r^{-2}$, the speed  of a fast blast wave decreases as $(t/\tau)^{-1/3}$, with $\tau$ a constant of proportionality that modulates the decreasing rate of the shock speed. Furthermore, \inlinecite{pinter1990} found that shock speed is convected by solar wind expansion. Thus, the speed ($v$) of an IP blast wave can be approximated by:
\begin{equation}
 \label{BWR_v}
  v =  \left( v_0 - w_{sw} \right) \left[ \frac{t}{\tau} \right]^{-1/3} + w_{sw} \, ,
\end{equation}
where $v_0$ is the initial shock speed (at $t=\tau$) and $w_{sw}$ the solar wind speed. We can express the leading front position of a blast wave ($r$) by integrating Equation (\ref{BWR_v}) over time:
\begin{equation}
 \label{BWR_r}
  r = w_{sw} \, t + \frac{3}{2} \tau^{1/3} \left( v_0 - w_{sw} \right) t^{2/3} - \frac{1}{2} \left( v_0 - w_{sw} \right) \tau + r_0 \, ,
\end{equation}
with $r_0$ being an integration constant related with the initial position of the shock front (at $t=0$). Notice that Equations (\ref{BWR_v}) and (\ref{BWR_r}) are valid for $t \ge \tau$. In order to simplify the integration of Equation (\ref{BWR_v}), we assumed $v=v_0$ for $0 \leq t < \tau$.

Then, from the relation between plasma frequency, solar wind density and heliocentric distance, we can define the position of source region by:
\begin{equation}
 \label{SF_r}
  r_i = r(t_i) = Q N \sqrt{n_{1AU}}\frac{1}{\underline{f}_i} \, .
\end{equation}
Where $n_{1AU}$ is the solar wind density at 1 AU and $Q=\sqrt{e^2 (1 {\rm AU})^2/ 4 \pi^2 \varepsilon_0 m_e }$, with $e$, $\varepsilon_0$, and $m_e$ being the fundamental charge, electric constant and the electron mass, respectively. Additionally, $N$ indicates the fundamental frequency ($=1$) or the first harmonic ($=2$) and $\underline{f}$ is  the lower frequency edge.

Now, by assuming that the location of TII source region is near the shock leading edge, we combine Equations (\ref{BWR_r}) and (\ref{SF_r}) to get:
\begin{equation}
 \label{BWR_relation}
  \frac{1}{\underline{f}_i} = a_3 t_i + a_2 t_i^{2/3} + a_0 \, .
\end{equation}
Equation (\ref{BWR_relation}) is a dispersion relation between frequencies and times that allows to reconstruct the kinematic evolution of IP shocks. The constants $a_3$, $a_2$ and $a_0$ are related with the blast wave equations by:
\begin{eqnarray}
w_{sw} &=& a_3 \sqrt{n_{1AU}} \, Q \, N \label{BWR_w} \, , \\
\tau &=& \left[ \frac{ a_3 r_0 - a_0 w_{sw} }{a_2 w_{sw}} \right]^{3/2} \label{BWR_tc} \, , \\
v_0&=& w_{sw} \left(1 + \frac{2 a_2}{3 a_3} \tau^{-1/3} \right) \label{BWR_v0} \, .
\end{eqnarray}
We calculate the values of $a_3$, $a_2$ and $a_0$ employing a gradient-expansion regression algorithm \cite{bevington2003}. Using the regression algorithm we can describe the kinematics of shocks by Equations (\ref{BWR_v}) and (\ref{BWR_r}), as well as  simulate the radio TII emission  by combining Equations (\ref{SF_r}) and (\ref{BWR_r}). In the Appendix we present the relations to estimate uncertainties associated to the BWR technique.

\section{Case events}
 \label{results}

We selected eight case events from the {\it WIND/WAVES type II bursts and CMEs} catalogue (\url{http://cdaw.gsfc.nasa.gov/CME_list/radio/waves_type2.html}) during the decaying phase of solar Cycle 23 (2000-2005). Table \ref{tabla_eventos} shows the event list. In all cases, the associated TII emissions required to be clearly identified and to last several hours. All events were catalogued as halo CMEs (Earth-directed), with (plane-of-sky) initial speeds above 900~$km s^{-1}$ (Table \ref{tabla_eventos}, column 3), and were associated with solar flares, as reported by SOHO LASCO CME \cite{gopalswamy2009}. The flares' locations were near the Sun's center (see last column of Table \ref{tabla_eventos}) to reduce possible directivity issues between the observer and TII source region. All the shocks and ICMEs were reported in \inlinecite{richardson2010} catalogue, but event~4 which possibly was missed due to a data gap in {\it ACE} registers (see \opencite{jian2006},  catalog). We identified event 4 from {\it WIND} {\it in situ} data. Column~6 in Table~\ref{tabla_eventos} shows the transit times (from near the Sun to 1~AU) of shocks and ICMEs. Column 7 shows the speeds of shocks and ICMEs at 1~AU obtained from {\it WIND} data. We estimated the shock local speed at 1~AU ($v_{1AU}$), applying velocity coplanarity on the upstream and downstream {\it in situ} data. We calculated the ICME speed averaging {\it WIND} {\it in situ} data on the ejecta leading edge. Since there was a data gap in event 5 impeding the use of the velocity coplanarity to estimate the shock local speed, for this case, we used instead as a proxy the shock transit speed to 1~AU.

The first step in the methodology was to study 1-min resolution TII frequency data following the procedure described in Figure \ref{figure0}. Subsequently, we reduced the frequency data temporal resolution to 0.5 hours, through an averaging process, in order to smooth out frequency irregularities to facilitate the convergence of the reconstruction algorithm. The averaging frequency data is presented in the bottom panels of Figures \ref{figure1} and \ref{figure2} as open-orange squares. Afterwards, we proceeded with the BWR technique by setting $r_0=6\, R_\odot$ (LASCO C2 external edge) for all the case events. The value of $n_{1AU}$ (see Table \ref{tabla_eventos}, column 4) was taken from {\it in situ} measurements 8-12 hours before the shock arrival. The initially supplied function for the regression algorithm used $v_0 = v_{cme0}$, $w_{sw}=w_{sw1}$ and $\tau = 5\,r_0/v_{cme0}$ (see Table \ref{tabla_eventos}). Additionally, we set $N=2$ for all the events (first harmonic frequency) and we used the detection time of each event as $t=0$ (Table \ref{tabla_eventos}, column 2).

\begin{table}
\rotatebox{90}{
\caption{List of events (chronograph and {\it in situ} measurements). From left to right: event number, date-time of CME detection, CME initial speed from LASCO observations; \textit{in situ} values for proton density and bulk speed of solar wind; shock (/ICME) transit times and shock (/ICME) local speed at 1~AU; and location (-class) of the associated solar flare.}
\label{tabla_eventos}
\begin{tabular}{cccccccc}     
  \hline                   
\multicolumn{2}{c}{Events}& \multicolumn{5}{c}{Event data}      & Flare \\
\# & Date-Hour$^a$    & ${v_{cme0}}^a$    & ${n_{1AU}}^b$ & ${w_{1AU}}^b$ & $TT_{sh}$/$TT_{cme}$ & ${v_{1AU}}^c$/${v_{cme1AU}}^d$ & Location$^e$-Class$^f$\\
   &   [yyyymmdd]-[UT]&[$km~s^{-1}$]  & [$cm^{-3}$]   &[$km~s^{-1}$]  &      [$h$]           & [$km~s^{-1}$]                  & \\
  \hline
1  & 20000606-15:54       & 1119  & 4.7 & 540 & 41.2/44.1 &  871/770  & N20E18-X2.3\\
2  & 20000714-10:54       & 1674  & 3.5 & 700 & 27.5/31.9 & 1120/1040 & N22W07-X5.7\\
3  & 20010426-12:30       & 1006  & 2.5 & 440 & 40.5/49.5 &  820/730  & N20W05-M1.5\\
4  & 20010924-10:31       & 2402  & 3.2 & 590 & 34.5/39.0 &  ~641/530$^b$& S16E23-X2.6\\
5  & 20011104-16:35       & 1810  & 7.5 & 420 & 33.5/34.9 & $^g$1206/750~~ & N06W18-X1.0\\
6  & 20011122-23:30       & 1437  & 5.5 & 419 & 31.5/38.5 & 1108/1040 & S17W36-M9.9\\
7  & 20040725-15:06       & 1333  & 1.8 & 599 & 31.9/35.1 & 1153/1000 & N08W33-M1.1\\
8  & 20050513-17:12       & 1689  & 3.5 & 410 & 33.5/36.9 & $^h$1111/950~~ & N12E11-M8.0\\
  \hline
\multicolumn {8}{l}{$^a$ {\it SOHO LASCO CME catalog} (\url{http://cdaw.gsfc.nasa.gov/CME_list/}).}\\
\multicolumn {8}{l}{$^b$ Detected \textit{in situ} by \textit{Wind} (\url{http://omniweb.gsfc.nasa.gov/}).}\\
\multicolumn {8}{l}{$^c$ Calculated by velocity coplanarity.}\\
\multicolumn {8}{l}{$^d$ Taken from \inlinecite{richardson2010} catalogue.}\\
\multicolumn {8}{l}{$^e$ SolarMonitor.org (\url{http://www.solarmonitor.org/}).}\\
\multicolumn {8}{l}{$^f$ NOAA classification.}\\
\multicolumn {8}{l}{$^g$ Estimated value $v_{1AU} = (1\,{\rm AU} - r_0)/TT_{sh}$ due data gap.}\\
\multicolumn {8}{l}{$^h$ Value taken from \inlinecite{bisi2010} due data gap.}
\end{tabular}
}
\end{table}

\subsection{Results}

 \begin{figure} 
 \centerline{\includegraphics[angle =90, width=0.8\textwidth,clip=]{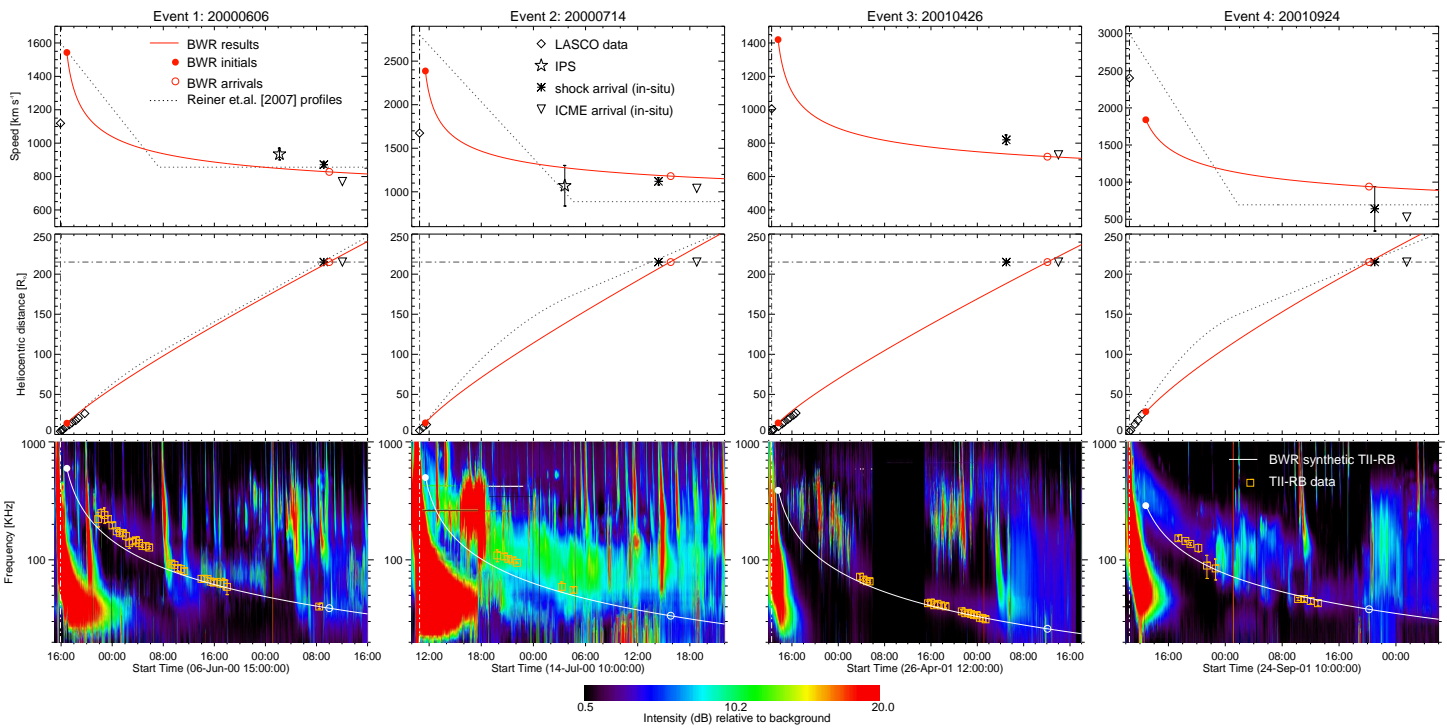}}
 \caption{Analysis results for events 1-4. Each column presents the results for one single event, while the rows present (from top to bottom) the results and data of speed, heliocentric distance and the dynamic radio spectrum, respectively. In upper and middle panels the open-diamonds, open-stars, asterisks and down-ward open-triangles represent different data. Due to space constrains, the references for the symbols and line styles in the two top rows are spread among the top panels. Solid-lines represent the shock trajectory obtained by the  BWR technique. The IPS velocity data is from STEL as reported by Gonzalez-Esparza and Aguilar-Rodriguez (2009). In the bottom panels, open orange-squares point out the averaging type II frequency data. See the text for further description. }\label{figure1}
 \end{figure}

\begin{figure} 
 \centerline{\includegraphics[angle =90, width=0.8\textwidth,clip=]{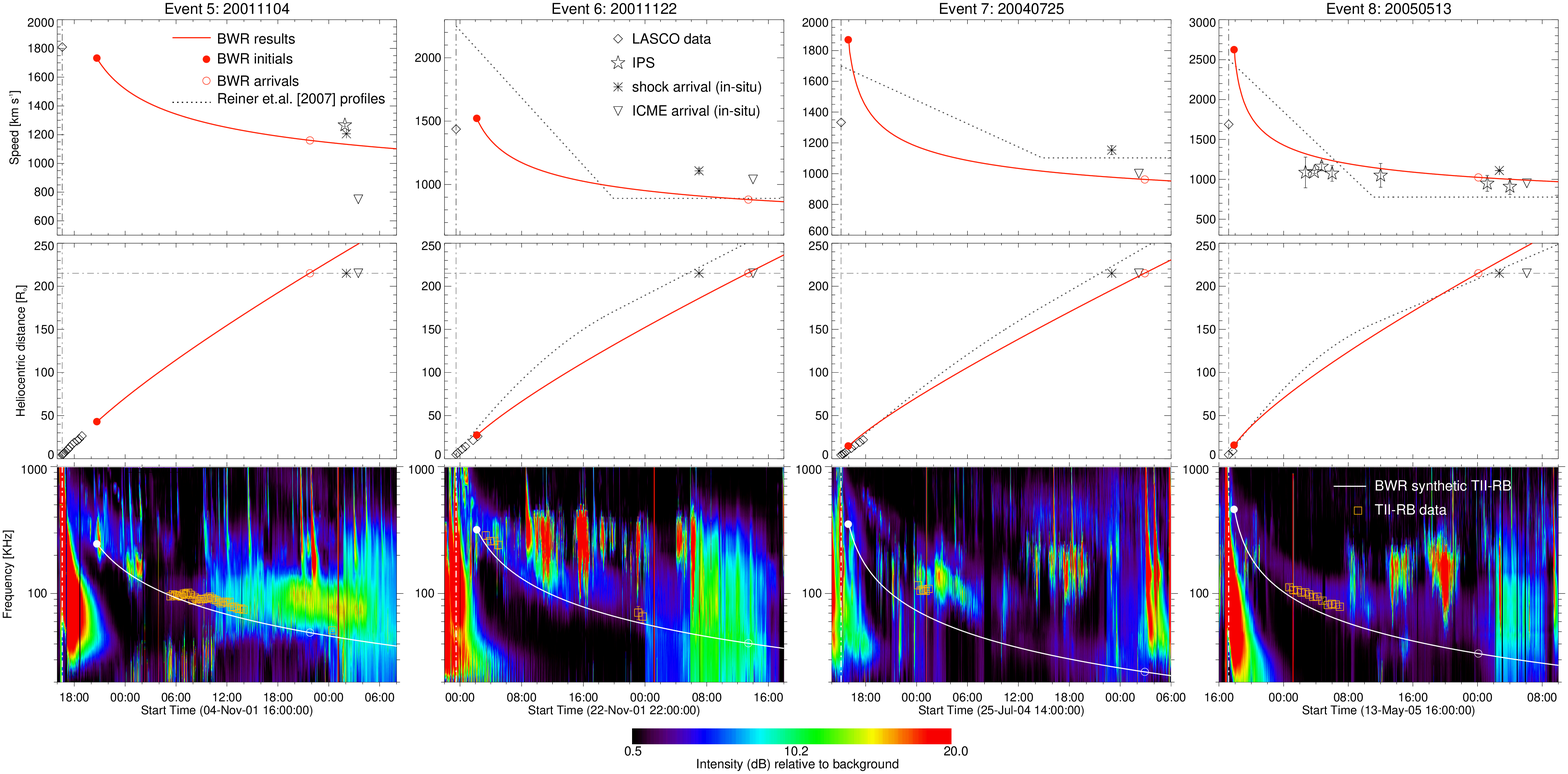}}
 \caption{Analysis results for events 5-8. Description is the same as Figure \ref{figure1}. The IPS velocity data is from STEL and ORT (2005/05/13) as reported by Gonzalez-Esparza and Aguilar-Rodriguez (2009) and Bisi {\it et.al.} (2010).}\label{figure2}
 \end{figure}

We present the results of the case events in Figures \ref{figure1} and \ref{figure2}. In both figures, each column represents a single event, meanwhile the rows, from top to bottom, show the speed and heliocentric distance of the shock fronts in time and the dynamic radio spectra. The vertical dash-dotted lines point out the event detection-time reported by LASCO CME catalogue (see Table \ref{tabla_eventos}, column 1). Upper and middle panels of the figures present the shock trajectories (speed and position versus time) obtained from the BWR technique  (solid-red lines). In the panels, we point out the initial (fill circles) and final (open circles) parameters calculated by the BWR technique. In these panels, for comparison with the BWR results, we also present data from: coronagraph images (open-diamonds) as reported by SOHO LASCO CME \cite{gopalswamy2009} catalog; {\it in situ} data of ICME (down-ward open-triangles) and shock (asterisks) arrivals at 1~AU; and IPS speed measurements (open-stars) as reported by \inlinecite{gonzalez2009} and \inlinecite{bisi2010}. In the bottom panels of the figures, we overplot the simulated radio emission (solid-white line) obtained from the BWR technique as well the radio data used in our analysis (open orange-squares).

In most cases, the simulated radio emission approximates very well the frequency data. For events 6 and 7 (Figure~\ref{figure2}) the limited TII data derived into poor BWR results. However, it is remarkable that, despite the limitation in the TII data, it was possible to apply the BWR technique in both events. Furthermore, in all cases the simulated frequency drift systemically limit the lower edge of the observed TII radio emission. 

\begin{table}
\rotatebox{90}{
\caption{Results from BWR analysis. From left to right: event number; values of $\tau$ and associated heliocentric distance ($r_\tau$), calculated value of the ambient solar wind ($w_{sw}$), calculated value of the shock initial speed ($v_0$); calculated shock transit time ($TT_{sh}$); calculated shock local speed at 1~AU  ($v_{1AU}$); proportional absolute difference between calculated and {\it in situ} registered values of $TT_{sh}$ and $v_{1AU}$; and angle ($\alpha$) between Sun-Earth line and location of associated solar flare. Column 7 also shows the direct difference between calculated and {\it in situ} registered values of $TT_{sh}$.}
\label{tabla_resultados}
\begin{tabular}{ccccccccc}     
  \hline                   
\multicolumn{1}{c}{Event} & \multicolumn{7}{c}{BWR results} & $\alpha$\\
\# & $\tau(r_\tau)^{a}$   & $w_0$           & $v_0$        & $TT_{sh}$ & $v_{1AU}$    & \multicolumn{2}{c}{Differences [$\%$]} &          \\
   & [$h$]([$R_\odot$])   & [$km~s^{-1}$]   & [$km~s^{-1}$]&   [$h$]   & [$km~s^{-1}$]& $TT_{sh}$([$h$])  &          $v_{1AU}$ & [$^\circ$]\\
  \hline
1        & 1.00(14.0) & 540 & 1542 & 42.1 & 828  & \,~2.2      (+0.9)&4.9       & 27\\
2        & 0.67(14.3) & 700 & 2387 & 28.9 & 1180 & \,~5.1      (+1.4)&5.4       & 23\\
3        & 1.11(14.1) & 440 & 1420 & 47.6 & 720  & 17.5        (+7.1)&12.2      & 21\\
4        & 2.30(27.9) & 317 & 1840 & 33.7 & 940  & \,~~2.3     (-0.8)&46.6      & 28\\
5        & 4.08(42.6) & 543 & 1733 & 29.2 & 1160 & $^b$12.8    (-4.3)&$^b$3.8   & 19\\
6        & 2.68(27.1) & 429 & 1522 & 37.9 & 881  & 20.3        (+6.4)&20.5      & 39\\
7        & 0.84(14.1) & 599 & 1870 & 35.8 & 962  & 12.8        (+4.1)&16.6      & 34\\
8        & 0.66(15.0) & 410 & 2625 & 30.9 & 1024 & \,~~7.8     (-2.6)&7.8       & 16\\
  \hline
Averages & 1.67(21.1) & 497 & 1867 & 35.8 & 908  & 10.0  ($^c$3.5)&13.7      & 26 \\
         &            &     &      &      &      &                & $^d$9.0  &    \\
  \hline
\multicolumn{9}{l}{$^a$ Heliocentric distance at $\tau$, $r_\tau = r_0+v_0\,\tau$.}\\
\multicolumn{9}{l}{$^b$ Estimated value due data gap.}\\
\multicolumn{9}{l}{$^c$ Absolute average.}\\
\multicolumn{9}{l}{$^d$ By neglecting event 4.}
\end{tabular}
}
\end{table}

In general, the resulting trajectories (position and speeds) are in good agreement with coronagraph, IPS and {\it in situ} measurements. Table \ref{tabla_resultados} summarizes the main results of the  analysis of the 8 case events, comparing the BWR calculations with coronograph and {\it in situ} data (Table~1). According to Equations (\ref{BWR_r}) and (\ref{BWR_v}), the calculated trajectory of the shock starts at $t=\tau$, time at which begins the shock blast wave propagation (Column 2). In general, the calculated shock initial positions ($t \sim \tau$) were located within the coronagraph (LASCO-C3) field of view. The calculated shock transit times (column~5) were quantitatively consistent with the {\it in situ} arrivals of the IP shocks. Column 7 lists the relative differences between calculated and {\it in situ} measurement of shock transit times ($TT$), where the average difference is about 10\%. Column 8 lists the relative differences between calculated and {\it in situ} measurement of shock speed at 1~AU ($v_{1AU}$). The average relative normalized difference was about 14\%. Note that the difference for event 4 is the largest; however, in this case, the estimation of the {\it in situ} shock speed has a large uncertainty. The fourth row of Figure~\ref{figure1} (upper panel) shows the largest error in the shock speed estimation. Neglecting event 4, the average associated difference of shock local speeds drops to 9\% (similar to the one for $TT_{sh}$).

The BWR calculations of the shock speed evolution in Figures \ref{figure1} and \ref{figure2} (upper panels) present an initial strong deceleration that rapidly decreases with time. Such deceleration considerably reduces ($\sim \, 50\%$) the speed of IP shocks when they arrive to 1 AU (see Table \ref{tabla_resultados}, columns 3 and 5). However, note that the deceleration is different for each event. We can estimate the decaying rate of shock speed by Equation (\ref{BWR_v}):
\[  \frac{d v}{dt} = - \frac{a_0}{3} \left( \frac{t}{\tau} \right)^{-3/4} \, , \]
where $a_0 = [v_0-w_{sw}]/\tau$ can be interpreted as an effective acceleration. Then, larger values of $a_0$, imply stronger decelerations acting on blast waves. Events 2 and 8 exemplify the effects of different values of $v_0-w_{sw}$ with similar $\tau$. For event 2 ($a_0=699\, m s^{-2}$) the decrease in speed at 1 AU was $\sim 50\%$; whereas for event 8 ($a_0=932\, m s^{-2}$) it was $\sim 61\%$. On the other hand, events 5 and 6 show the case for different values of $\tau$ and close values of $v_0-w_{sw}$. For event 5 ($a_0=81\, m s^{-2}$) the shock speed decreases $\sim 33\%$; whereas in event 6 ($a_0=113\, m s^{-2}$) the reduction was $\sim 42\%$.

The BWR calculated shock parameters at 1~AU for events 3, 5, 6 and 7 present differences with {\it in situ} data. Such differences may derive from contamination of the TII radio spectrum and/or directivity issues. On one hand, events 3 and 5 occurred during high solar activity intervals with occurrence of other CMEs before and after. This could lead into multiple (complex) CMEs propagating through an unstable ambient solar wind. This scenario might contaminate the TII dynamic radio spectra, misleading the results. On the other hand, the associated-flare locations of events 6 and 7 have a large angular distance ($\sim 35^\circ$) with respect to the Sun-Earth line (see Table \ref{tabla_resultados}, column 8). Using the flare location as a proxy for the CME propagation direction, we can estimate if the source emission is propagating far away from the Sun-Earth line. For these cases, the radio source may not be directly seen by the observer. TII emissions are affected by the line of sight between the observer and the source \cite{knock2005}. This might be the case in event 7, where the TII emission does not appear as a continuous frequency drift, but like a chain of consecutive bursts. For events 6 and 7, this directivity issue might explain the reduced number of TII radio data, resulting into poor BWR results. In Figure \ref{figure2} the simulated TII emission of events 6 and 7 barely follows the TII spectral data.

\section{Discussion}
 \label{discussion}

We present the BWR technique to reconstruct the evolution of shock fronts using TII emission data. The technique is based on three main assumptions: (1) solar wind density decays as $r^{-2}$ and expands with constant speed ($w_{sw}$); (2) the TII emission source is located somewhere near the leading edge of the shock front; and (3) IP shocks evolve as blast waves. According to the first assumption, this technique would be more suitable for quiet periods of solar activity, when isolated fast CME events propagate through a quasi-stationary ambient solar wind. On the other hand, note that the average initial positions of blast wave propagation ($r_\tau$), given by the BWR results, were located around $21\,R_\odot$ (see Table \ref{tabla_resultados} column 2) which is beyond the supermagnetosonic point.

With respect to the second assumption, as some authors suggest, the source region of TII emissions might be located far from the shock's leading edge \cite{knock2003}. This source location may affect the results of this technique. The two events in the list whose associated flare locations were more than $30^\circ$ from the Sun center (events 6 and 7 in the last column of Table~2) had significant differences with respect to the {\it in situ} measurements. It is possible that these errors were related to the location of the source regions, and that the BWR technique is more suitable for Earth-directed events, where source regions are close to the Sun-Earth line.

It is not clear where ICMEs get exhausted of driving IP shocks in the inner heliosphere. However, beyond that critical distance ($r_\tau$), the IP shocks should evolve as blast waves, as the third assumption dictates. As commented in the Introduction, there are multiple works that suggest that IP shocks evolve as blast waves. For example, \inlinecite{burlaga1981}, by analyzing ICMEs/shocks detected by Voyager 1 and 2, found that the shock might be no longer driven by ICMEs at heliocentric distances around 2~AU. Departing from geometrical analysis, \inlinecite{feng2010} found that at least $34\%$ of shocks detected {\it in situ} (at 1 AU) were not driven by their associated ICMEs. \inlinecite{corona2011} studied the momentum fluxes between ICMEs and plasma sheaths using numerical simulations and concluded that IP shocks are no longer driven by ICMEs when they reach 1 AU. The results from our study suggest that IP shocks evolve as blast waves beyond $\sim 21\,R_\odot$ ($t>\tau$). This value is consistent with the ones separately found by \inlinecite{pinter1990} ($\sim 0.12\, {\rm AU} = 25.8\, R_\odot$) and \inlinecite{corona2012c} ($\sim 44.7\, R_\odot$) through empirical and analytic approaches, respectively. Furthermore, by assuming blast wave propagation for IP shocks, the BWR technique was able to reconstruct speed and position profiles that showed qualitative and quantitative agreements with a number of different data. Furthermore, blast wave propagation explains the well known fact of deceleration of IP shocks, as well as the unexpected growth in stand-off distances found by \inlinecite{maloney2011}.

In Figures 2 and 3 (upper and middle panels) we plot the shock trajectories using the approximation by \inlinecite{reiner2007} (dotted grey line). They proposed a two stages kinematic model to proxy the trajectory of fast ICMEs/shocks. Such a model consists in a pristine period of intense deceleration followed by a constant speed stage. Then they search for the trajectories of ICME/shocks that simultaneously satisfied data from coronagraph images, {\it in situ} registers and TII emissions. Comparing our solutions with the ones obtained by \inlinecite{reiner2007}, we notice similarities between them. Blast wave speeds present an initial strong deceleration which rapidly decreases with time. This behavior resembles the one presented by the two stages kinematic model. Event 1 is an example where there is a good agreement between the two methodologies which both predict very similar shock arrival times to 1 AU. 

Though BWR technique is presented as a tracking method for IP shocks; it might be useful for forecasting proposes. According to our results, BWR is more suitable, but not restricted, for Earth-directed shocks occurring during quiet periods of solar activity. Our calculated shock trajectories showed a tendency to arrive later than their {\it in situ} counterparts. The average difference between calculated and {\it in situ}-registered transit times was 3.5 h, a value similar to others ($\sim 7\,{\rm h}$) derived from empirical ({\it e.g.} \opencite{cremades2007}; \opencite{feng2009}; \opencite{mostl2011}) and analytical ({\it e.g.} \opencite{feng2006}; \opencite{song2010}) arrival-predicting tools. An important limitation of BWR technique, for predicting proposes, is the acquisition of {\it in situ} data. This data is taken around 8-12 hours before shock arrival, leaving few hours to issue alarms. Though this limitation can be easily solved by using IPS measurements or average values of solar wind density at 1 AU, exploring this topic escapes the scopes of this work.

The BWR calculated shock speed profiles were consistent with coronagraph, {\it in situ} and IPS data. However, although our results showed good agreement with different data, we have to point out that those data correspond to different but related structures. On one hand, coronagraph images give information about (plane-of-sky) propagation of CMEs in the solar corona. On the other hand, IPS speeds might be related to plasma sheaths behind the IP shocks, and the {\it in situ} measurements to local characteristics of ICMEs and shocks at 1 AU. 

The BWR technique is sensitive to contamination and directivity issues on TII emission data. The applicability of BWR, and the certainty on its results, will depend on the availability and quality of the TII data. However, the calculated transit times and arrival speeds of shocks were quantitatively consistent with their {\it in situ} registered counterparts. In average, our results showed $\sim 90\%$ of agreement with {\it in situ} data (see Table \ref{tabla_resultados}, column 7). Even for unfavorable events (say events 3, 5, 6, 7), it was possible to approximate $TT_{sh}$ and $v_{1AU}$ with reasonable agreement with the {\it in situ} data.


\section{Conclusions}
 \label{conclusions}
 
We introduced the blast wave regression (BWR) technique. This technique reconstructs the evolution of fast ICMEs/shocks associated with type II radio bursts.  The technique assumes that interplanetary shocks propagate as blast waves through a  stationary homogeneous ambient solar wind, whose density decays with the square of the heliocentric distance. Using frequency  data from the TII emission, the technique develops a dispersion relation whose regression solves the kinematics (speed and position) of the IP shock. We applied the BWR technique to analyze 8 ICME/shock events associated with TII emissions occurred during the decaying phase of solar cycle 23. The results suggest that the shock events propagated as blast waves beyond $\sim 21\,R_\odot$. The BWR results also showed notable consistencies with data from coronagraph images, speeds deduced from IPS observations and {\it in situ} measurements. The BWR calculated transit times and arrival speeds (at 1~AU) of shocks showed about $90\%$ of consistency with the {\it in situ} data. The accuracy of this technique depends on the availability and quality of the type II spectral data associated with the ICME/shock. The results may be mislead by contamination of radio data and directivity issues. However, even for those unfavorable conditions, the technique was capable to estimate transit times and arrival speeds that differed around $20\%$ from their {\it in situ} counterparts. The agreement that our results showed with different data sets, preliminarily indicates that the BWR technique could be a useful tool for ICME/shock tracking and arrival forecasting, although the performance analysis of the latter is not under the scope of this paper.

%

%

%
 \appendix   
  \label{apendice}

\section*{Uncertainties} 

We can estimate the uncertainties associated with the BWR technique by using the differentiation method. Then, the uncertainty ($\delta r_i$) of $r_i$ is given by:
\begin{equation}
 \label{SF_dr}
  \frac{\delta r_i}{r_i} = \frac{\delta n_{1AU}}{n_{1AU}} + \frac{\delta \underline{f}_i}{\underline{f}_i}\, ,
\end{equation}
where $\delta n_{1AU}$ and $\underline{f}_i$ are the uncertainties of $n_{1AU}$ and $\underline{f}_i$, respectively. Here we can estimate $\delta \underline{f}_i \sim \Delta f_i/4$. Finally, the uncertainties on shock trajectory are expressed by:
\begin{eqnarray}
\frac{\delta w_{sw}}{w_{sw}} &=& \frac{\delta n_{1AU}}{2 n_{1AU}}  \, , \label{BWR_dw}\\
\frac{\delta \tau}{\tau} &=& \frac{3}{2} \left[ \frac{ a_3 \delta r_0 - a_0 \delta w_{sw} }{a_3 r_0 - a_0 w_{sw}} + \frac{\delta w_{sw}}{w_{sw}}\right] \label{BWR_dtc} \; {\rm and} \\
\frac{\delta v_0}{v_0}       &=& \frac{\delta w_{sw}}{w_{sw}} + \left[ \frac{2 a_2}{ 9 a_3 \tau^{1/3} + 6 a_2 } \right] \frac{\delta \tau}{\tau} \label{BWR_dv0} \, .
\end{eqnarray}
Where $\delta w_{sw}$, $\delta \tau$ and $\delta v_0$ are the errors of solar wind speed, $\tau$ and initial speed of shock front, respectively. Additionally, $\delta n_{1AU}$ is the uncertainty related to $n_{1AU}$ and $\delta r_0$ the uncertainty of $r_0$.

%
 \begin{acks}
P. Corona-Romero, V. De-la-Luz and J.C. Mejia-Ambriz thanks Catedras-CONACyT project 1045: \textit{Servicio de Clima Espacial Mexico} (SCiESMEX). J. A. Gonzalez.-Esparza acknowledges the partial funding by the DGAPA-PAPIIT grant IN109413 and CONACyT grant 152471. E. Aguilar-Rodriguez thanks DGAPA-PAPIIT grant IN103615, and CONACyT grant 220981.
 \end{acks}

%
%
 \bibliographystyle{spr-mp-sola}
 \bibliography{referencias}  
%
%
%
%

\end{article} 
\end{document}